\documentclass[prl,aps,twocolumn,superscriptaddress,nofootinbib,longbibliography]{revtex4-2}
\usepackage{graphicx} 
\usepackage{dcolumn}  
\usepackage{bm}       
\usepackage{amsmath}
\usepackage{epsfig}
\usepackage{color}
\usepackage{ulem}

\begin{document}
\title{Thermal radiation forces on planar structures with asymmetric optical response}
\author{Juan R. Deop-Ruano}
\affiliation{Instituto de \'Optica (IO-CSIC), Consejo Superior de Investigaciones Cient\'ificas, 28006 Madrid, Spain}
\author{F. Javier Garc\'ia de Abajo}
\email{javier.garciadeabajo@nanophotonics.es}
\affiliation{ICFO-Institut de Ciencies Fotoniques, The Barcelona Institute of Science and Technology, 08860 Castelldefels, Barcelona, Spain}
\affiliation{ICREA-Instituci\'o Catalana de Recerca i Estudis Avançats, Passeig Llu\'is Companys 23, 08010, Barcelona, Spain}
\author{Alejandro Manjavacas}
\email{a.manjavacas@csic.es}
\affiliation{Instituto de \'Optica (IO-CSIC), Consejo Superior de Investigaciones Cient\'ificas, 28006 Madrid, Spain}

\date{\today}

\begin{abstract}
Light carries momentum and, upon interaction with material structures, can exert forces on them. Here, we show that a planar structure with asymmetric optical response is spontaneously accelerated when placed in an environment at a different temperature. This phenomenon originates from the imbalance in the exchange rates of photons between both sides of the structure and the environment. Using a simple theoretical model, we calculate the force acting on the planar structure and its terminal velocity in vacuum, and analyze their dependence on the initial temperature and the geometrical properties of the system for different realistic materials. Our results unravel an alternative approach to manipulating objects in the nano and microscale that does not require an external source of radiation.
\end{abstract}
\maketitle

Optical forces are widely used to manipulate material structures \cite{ZSZ22}. A paradigmatic example is optical tweezers \cite{A1970,ADY1987,G03,JRQ11, BCL21,VMR23}, which enable the precise trapping and displacement of objects as diverse as nanoparticles \cite{MJG13} and live cells \cite{ZWZ13_2}. In the macroscale, optical forces have been studied in the context of space propulsion \cite{LBM18}. In all of these scenarios, the optical force is exerted by an externally applied light source.   
However, even in the absence of an external source, the fluctuations of the electromagnetic field, originating from its quantum nature and the population of photonic modes at finite temperature, can give rise to forces and torques \cite{L07,ama7,RCJ11,GGL15,ama50,SGP18,ama66,ACG20,XJG22,SGM22} on material structures. These are usually referred to as Casimir interactions \cite{C1948,DMR11,WDT16,RIB22} and offer an alternative approach to manipulate nano and microstructures. For instance, they can be exploited to achieve levitation \cite{LMR10,ECH15, LZ16} and build self-assembled cavities \cite{MCK21}. 

The fluctuations of the electromagnetic field are also at the origin of radiative heat transfer between objects at different temperatures \cite{NSH09,RSJ09,SGZ14,KSF15,BMF16,CG18}, which is the only mechanism responsible for their thermalization in vacuum \cite{ama75}. The interplay between radiative heat transfer and Casimir interactions gives rise to interesting phenomena. These include, among others, the possibility of reversing heat transfer between rotating objects \cite{ama85}, as well as thermal friction acting on objects moving with respect to an environment at finite temperature \cite{MPP03, V15, DK14}. 

Interestingly, the inverse process (\textit{i.e.}, the use of thermal radiation to propel structures) has remained mostly overlooked. Recently, it has been proposed that Janus particles placed in an environment at a different temperature can achieve self-propulsion \cite{RMP17}. This type of structure has two distinct sides made of different materials \cite{ZFD21}. As a consequence, the physical and chemical characteristics of these objects display a strong asymmetry, which is being exploited in a variety of applications including biosensing \cite{HVL22}, color switching \cite{NKH20}, and photocatalysis \cite{WLL18}, as well as to achieve highly precise optical nano and micromanipulation \cite{NCK15, IKL16,BCG23}.

Here, we analyze the optical force exerted by thermal radiation on a planar structure made of two layers of different materials. The simplicity of the system allows us to derive analytical expressions that characterize the transfer of energy and momentum between the structure and the environment. In doing so, we calculate the force and acceleration for different combinations of realistic materials, thicknesses, and temperatures. Furthermore, we compute the temporal evolution of the temperature and velocity, thus fully characterizing the motion of the structure. The dynamics of our system resembles the movement of Janus structures in liquid solutions due to self-thermophoresis \cite{JYS10, BSM13, QPG17}, in which the force originates from a local thermal gradient. However, in our system, the motion is solely induced by the different rates of exchange of photons between the environment and the two sides of the structure. Therefore, it does not require the presence of a medium nor the use of an external source of light. The results of our work serve to advance our understanding of the optical forces produced by thermal radiation, thus helping to leverage them for the manipulation of nano and microstructures. 

\begin{figure}
\begin{center}
\includegraphics[width=80mm,angle=0]{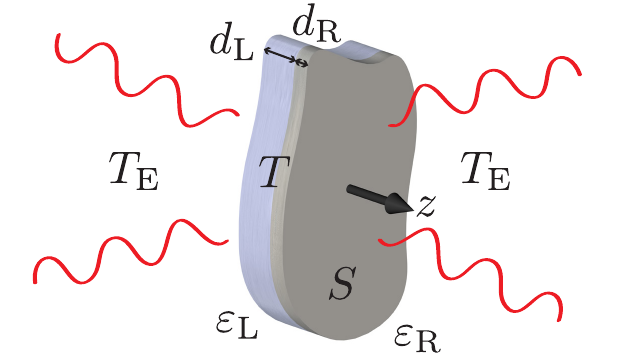}
\caption{Schematics of the system under study, which consists of a planar structure composed of two layers with thicknesses $d_{\rm L}$ and $d_{\rm R}$, made of different materials with dielectric permittivities $\varepsilon_{\rm L}$ and $\varepsilon_{\rm R}$. Both layers have a surface area $S$ and share the same temperature $T$.  The structure is surrounded by an environment at $T_{\rm E}=300\,$K.} \label{fig1}
\end{center}
\end{figure}

The system under consideration is depicted in Fig.~\ref{fig1}. It consists of a planar structure composed of two (left and right) layers made of different materials with dielectric permittivities $\varepsilon_{\rm L}$ and $\varepsilon_{\rm R}$. The left and right layers have thickness $d_{\rm L}$ and $d_{\rm R}$, respectively. The surface area of both layers is $S$ and they share a temperature $T$. The temperature of the environment is $T_{\rm E} = 300\,$K throughout this work. We assume that the lateral dimensions are much larger than $d_{\rm L}$, $d_{\rm R}$, and the thermal wavelength $\lambda_{T}=2\pi\hbar c/(k_{\rm B}T)$, which allows us to treat the layers as infinite thin films.

\begin{figure}
\begin{center}
\includegraphics[width=80mm,angle=0]{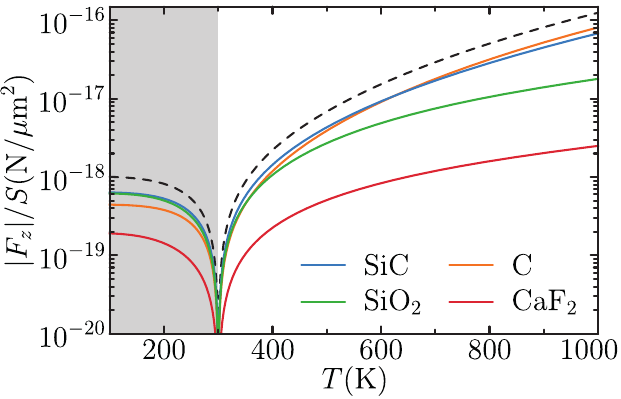}
\caption{Force per unit area exerted on the planar structure as a function of $T$. We assume the structure of Fig.~\ref{fig1} with a left (absorbing) layer of thickness $d_{\rm L}=10\,\mu$m made of one of the materials indicated by the legend and a right (reflecting) Ag layer of thickness $d_{\rm R}=0.1\,\mu$m. The dashed black curve represents the upper limit for the force given by Eq.~(\ref{eq_Fi}). The shaded area indicates the range of $T$ for which the force is negative. The environment is at temperature $T_{\rm E}=300\,$K.} \label{fig2}
\end{center}
\end{figure}

The optical response of the planar structure is determined by its reflectance and absorbance, which we denote $\mathcal{R}^{\nu}_i(\omega, \mu)$ and $\mathcal{A}^{\nu}_i(\omega, \mu)$, respectively. These quantities depend on the frequency $\omega$ and $\mu = \cos(\theta)$, where $\theta$ is the polar angle measured with respect to the surface normal. We use the superscript $\nu=s,p$ to indicate the polarization and the subscript $i={\rm L},{\rm R}$ to distinguish the properties of the left and right sides of the structure. The net exchanged power per unit area by the structure with the environment can be calculated as 
\begin{equation}
\begin{aligned}
\frac{P}{S}=\int_0^{1}  \! \!  \mu d\mu   \int_0^{\infty} \! \! d \omega \Delta u (\omega,T,T_{\rm E}) \bar{\mathcal{A}}(\omega, \mu), \label{eq_P}
\end{aligned}
\end{equation}
where $\Delta u (\omega,T,T_{\rm E}) = u (\omega, T) - u (\omega, T_{\rm E})$ and $u(\omega, T) ={\hbar}\omega^3 n\left(\omega, T\right)/({4 \pi^2 c^2})$, with $n(\omega, T) = [{\rm exp}(\hbar \omega / k_{\rm B}  T) - 1]^{-1}$ being the Bose-Einstein distribution. Furthermore, $\bar{\mathcal{A}}(\omega,\mu)= \sum_{\nu}\left[ \mathcal{A}^{\nu}_{\rm L} (\omega,\mu)+\mathcal{A}^{\nu}_{\rm R} (\omega,\mu)\right]$. 
To calculate the force exerted on the structure by the radiation exchanged with the environment, we need to compute the momentum transferred by the photons absorbed, emitted, and reflected. Specifically, each photon absorbed or emitted by the system transfers a momentum along the $z$-axis equal to $\pm\mu \hbar\omega/c$, with the upper (lower) sign corresponding to the left (right) layer. This value doubles for each (specularly) reflected photon. 
Adding all of these contributions, the force per unit area reads
\begin{equation}
\begin{aligned}
\frac{F_z}{S}= \frac1{c}\int_0^{1}  \! \! \mu^2 d\mu   \int_0^{\infty} \! \! d \omega  \Delta u (\omega,T,T_{\rm E})\Delta \mathcal{R}(\omega, \mu), \label{eq_F} 
\end{aligned}
\end{equation}
with $\Delta \mathcal{R} (\omega,\mu) =  \sum_{\nu} \left[ \mathcal{R}^{\nu}_{\rm R}(\omega, \mu)- \mathcal{R}^{\nu}_{\rm L}(\omega, \mu)\right]$.
Upon inspecting Eq.~(\ref{eq_F}), it becomes clear that the upper limit for the force is reached by a structure that, for all values of $\omega$ and $\mu$, is a perfect absorber on one side and a perfect reflector on the other side. In such a limit, we can analytically solve the integrals over frequencies and angles in Eqs.~(\ref{eq_P}) and (\ref{eq_F}). In particular, choosing $\mathcal{A}_{\rm L}(\omega, \mu) = 1$ and $\mathcal{R}_{\rm R}(\omega, \mu) = 1$, the net exchanged power between the system and the environment becomes
\begin{equation}
\frac{P}{S}=\sigma_{\rm SB}\left(T^4-T_{\rm E}^4\right), \nonumber 
\end{equation}
while the force reduces to
\begin{equation}
\frac{F_z}{S}=\frac{2}{3 c}\sigma_{\rm SB}\left(T^4-T_{\rm E}^4\right)  \label{eq_Fi} 
\end{equation}
where $\sigma_{\rm SB} = \pi^2 k_{\rm B}^4/(60 c^2 \hbar^3)$ is the Stefan-Boltzmann constant. Indeed, the net exchanged power, which depends on the fourth power of the temperatures, coincides exactly with the Stefan-Boltzmann law. This is an expected result because one of the sides of the structure behaves as a blackbody, while the other one does not exchange any power. The force, given by Eq.~(\ref{eq_Fi}), is proportional to the net exchanged power and satisfies the simple relation $F_z = 2P/(3c)$, where the $1/c$ factor is the constant of proportionality between the energy and momentum of a photon, while the $2/3$ factor accounts for the angular averaging of the force component along the $z$-axis. 
The dashed black curve in Fig.~\ref{fig2} shows the outcome of Eq.~(\ref{eq_Fi}) as a function of the temperature of the system. For the range of temperatures under consideration, the force per unit area reaches values up to $10^{-16}\,$N$/\mu$m$^2$. As expected, the sign of the force is positive when $T>T_{\rm E}$ and negative in the opposite case (shaded area).

This result shows that the thermal radiation exchanged with the environment exerts a substantial force on a planar structure composed of absorbing and reflecting layers. To analyze whether such a phenomenon is experimentally feasible, in the remainder of this work, we consider systems made of realistic materials. Specifically, we choose Ag for the layer acting as a reflector. This material behaves as an almost perfect conductor for frequencies below the visible range. As a consequence, we consider a Ag layer with a thickness of $d_{\rm R}=0.1\,\mu$m, which results in $\mathcal{R}_{\rm R}(\omega, \mu) \approx 1$ for the relevant frequency range. Regarding the layer acting as an absorber, we need a material displaying a strong absorption in the infrared regime, where $\Delta u (\omega,T,T_{\rm E})$ takes significant values. To that end, we consider layers made of amorphous C, SiC, SiO$_2$, and crystalline CaF$_2$. Using tabulated dielectric functions for these materials compiled in Refs.~\cite{HGK75,LPG11,RLM16_2, KPG12,KPJ07,M1963,KSK1962}, and a Drude function for Ag with parameters taken from Ref.~\cite{YDS15}, we calculate $F_z/S$ by numerically integrating Eq.~(\ref{eq_F}).  We begin by assuming a thickness $d_{\rm L} = 10\,\mu$m for the absorbing layer but also discuss below the dependence of the force on this parameter. The corresponding results are displayed in Fig.~\ref{fig2} (solid curves). We observe that, except for CaF$_2$, the obtained forces are of the same order of magnitude as the ideal system discussed above. 
Depending on the temperature of the structure, the largest values of the force are reached by the system with SiC or C. This behavior is the direct consequence of the different spectral overlap between $\Delta u(\omega, T, T_{\rm E})$ and $\Delta \mathcal{R}(\omega,\mu)$ for these materials. 

\begin{figure}
\begin{center}
\includegraphics[width=80mm,angle=0]{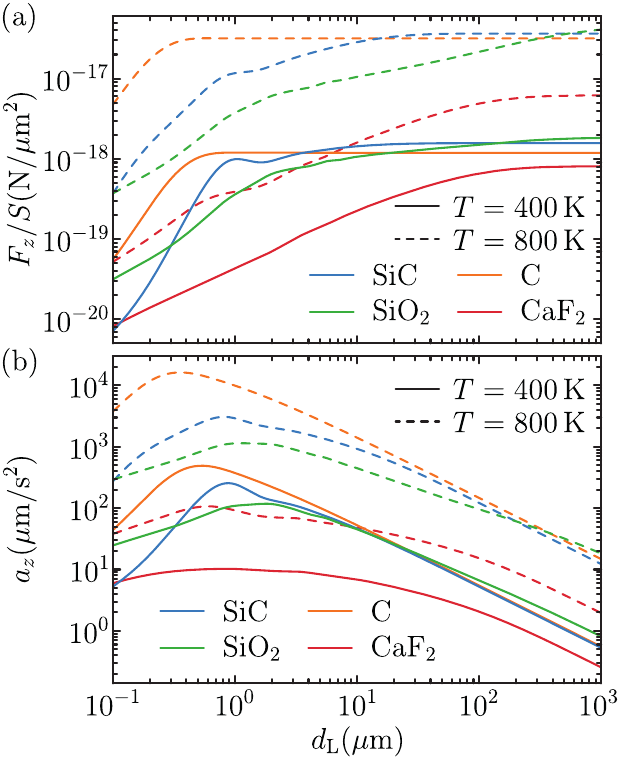}
\caption{Force per unit area (a) and acceleration (b) of the planar structure as a function of $d_{\rm L}$. Solid and dashed curves correspond to structure temperatures $T = 400\,$K and $T = 800\,$K, respectively, while the environment is at $T_{\rm E}=300\,$K. We show results for different materials in the left layer (see legend) and a right Ag layer of thickness $d_{\rm R}=0.1\,\mu$m in all cases.} \label{fig3}
\end{center}
\end{figure}

The thickness of the absorbing layer plays a crucial role in the force, as well as in the resulting acceleration of the planar structure. The absorbance, and hence the force, is expected to increase with $d_{\rm L}$ until a saturation value is reached, once the material becomes optically thick for the relevant frequencies. However, a thicker layer has a larger mass and, therefore, a smaller acceleration. In Figs.~\ref{fig3}(a) and (b), we analyze the force per unit area and the resulting acceleration as a function of $d_{\rm L}$ for $T=400\,$K (solid curves) and $T=800\,$K (dashed curves). As anticipated, the force increases monotonically with thickness, except for the structure with SiC at $T=400\,$K, which displays a resonant behavior around $1\,\mu$m produced by multiple reflections at the interfaces. Furthermore, the thickness for which the force saturates depends strongly on the material, ranging from $\lesssim 1\,\mu$m for C to $\approx 1000\,\mu$m for SiO$_2$. These differences result in a nontrivial behavior for the acceleration. This quantity, which is analyzed in Fig.~\ref{fig3}(b), is derived from the force as $a_z = F_z / m$, with $m = S (\rho_{\rm L} d_{\rm L} + \rho_{\rm R} d_{\rm R})$ being the mass of the structure, while $\rho_{\rm L}$ and $\rho_{\rm R}$ are the mass densities of the two layers. Here, we use $10490$, $2150$, $2980$, $2250$, and $3180\,$kg/m$^3$ for the mass density of Ag, C, SiC, SiO$_2$, and CaF$_2$. Analyzing the results for the acceleration, we observe that the trade-off between the force and the mass results in an optimum value of $d_{\rm L}$ at which the acceleration is maximized. Interestingly, the structure with C reaches the largest acceleration for both temperatures under consideration: $\approx 490\,\mu$m/s$^{2}$ for $d_{\rm L}\approx0.53\,\mu$m and $T=400\,$K, and $\approx 16280\,\mu$m/s$^{2}$ for $d_{\rm L}\approx0.36\,\mu$m and $T=800\,$K. This behavior is consistent with the fact that this structure displayed the largest force for small thicknesses. The systems with SiC and SiO$_2$ also reach maximum accelerations of approximately the same order of magnitude but the values for CaF$_2$ are significantly smaller.

\begin{figure}
\begin{center}
\includegraphics[width=80mm,angle=0]{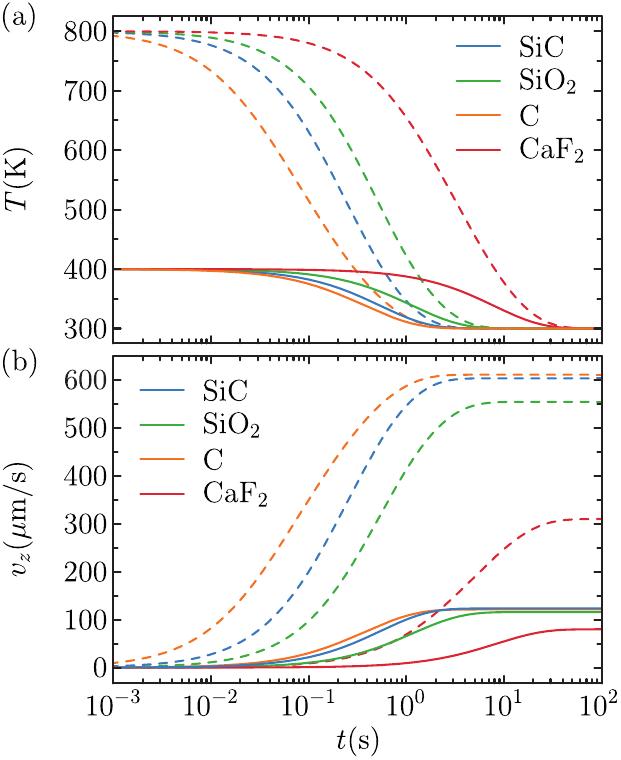}
\caption{Temporal evolution of the temperature $T(t)$ (a) and velocity $v_z(t)$ (b) of the planar structure. Solid and dashed curves show results for an initial temperature of $T(0) = 400\,$K and $T(0) = 800\,$K, respectively, while the initial velocity is $v_z(0) = 0$ in all cases. We present results for the different materials in the left layer (see legend), with the right layer being always made of Ag. We set $d_{\rm L}= 1\,\mu$m, $d_{\rm R}= 0.1\,\mu$m, and $T_{\rm E}=300\,$K.} \label{fig4}
\end{center}
\end{figure}

The force experienced by the planar system originates in the exchange of photons with the environment, which, in turn,  is fueled by the temperature difference between the system and the environment. Therefore, as the structure accelerates, it thermalizes with the environment and eventually reaches a terminal velocity. To analyze this process, we consider the rate of change of the internal energy of the system, which is given by $\dot{U} = -P - F_zv_z$  with $v_z$ being the velocity of the structure. Since, as we have discussed above, the force can be approximated as $F_z \approx  2P / (3c)$ and $v_z/c \ll 1$, it follows that $\dot{U} \approx -P$. Furthermore, since the conduction of heat across the system occurs over a much shorter time scale than thermalization with the environment, it is reasonable to assume that both layers of the structure effectively remain at the same temperature. With all of these assumptions, the rate of change of the temperature of the system reads $\dot{T} = -P/\Gamma$, where $\Gamma = S(\gamma_{\rm L} \rho_{\rm L} d_{\rm L} + \gamma_{\rm R} \rho_{\rm R} d_{\rm R}) $, while $\gamma_{\rm L}$ and $\gamma_{\rm R}$ are the specific heat capacities of the two layers. Then, the temporal evolution of the temperature of the planar system is given by
\begin{equation}
\begin{aligned}
T(t) = T(0)-\frac{1}{\Gamma} \int_0^{t} d t^{\prime} P(t^{\prime}). \nonumber
\end{aligned}
\end{equation}
Fig.~\ref{fig4}(a) shows the temperature of the planar structure as a function of time for the initial conditions $T(0)=400\,$K (solid curves) and $T(0)=800\,$K (dashed curves). We assume $d_{\rm L}= 1\,\mu$m and use $235$, $721$, $671$, $740$, and $881\,$J/(kg\,K) for the specific heat capacities of Ag, C, SiC, SiO$_2$, and CaF$_2$.
As in Figs.~\ref{fig2} and ~\ref{fig3}, the different colored curves are the results corresponding to the different materials under consideration. As expected, the temperature of the systems evolves towards $T_{\rm E}$ in all cases. However, while for the planar structures with C, SiC, SiO$_2$ the thermalization occurs in a time scale of seconds, for CaF$_2$, this process takes an order of magnitude more time. 

As thermalization occurs, the system is accelerated, and its velocity, which can be calculated as 
\begin{equation}
\begin{aligned}
v_z(t) = v_z(0)+ \frac{1}{m} \int_0^{t} d t^{\prime} F_z(t^{\prime}), \nonumber
\end{aligned}
\end{equation}
is plotted in Fig.~\ref{fig4}(b) assuming that $v_z(0)=0$. As anticipated, the velocity grows until it reaches a terminal value at the time when the system thermalizes with the environment. For the smallest initial temperature under consideration, the terminal velocity of the different materials here analyzed lies in the range from $80\,\mu$m/s for CaF$_2$ to $120\,\mu$m/s for C and SiC. These values become approximately five times larger for $T(0)=800\,$K. 
Interestingly, while the system with C displays the largest acceleration for this value of $d_{\rm L}$, its terminal velocity is similar to that reached by the systems with SiC and SiO$_2$. 
The reason is that, as shown in Fig.~\ref{fig4}(a), the thermalization process for the system with C is the fastest.

During the acceleration, the system transforms thermal energy into kinetic energy. It is interesting to estimate the efficiency of this process. To that end, we define $\eta=K/E$ as the ratio between the final kinetic energy and the initial thermal energy stored in the system. The former is given by $K=mv_{z,\infty}^2/2$, with $v_{z,\infty}$ being the terminal velocity, while the latter reads $E=\Gamma|T(0)-T_{\rm E}|$. For the ideal system described by Eq.~(\ref{eq_Fi}), the terminal velocity admits the simple expression 
\begin{equation}
v_{z,\infty} = \frac{2}{3 c}\frac{\Gamma}{m}[T(0) - T_{\rm E}], \nonumber
\end{equation} 
which allows us to obtain the following estimate: $\eta = |v_{z,\infty}|/(3 c)$. Therefore, for a fixed initial temperature difference, achieving a higher value of $\eta$ requires a large value of $\Gamma$ (so the structure thermalizes with the environment at a slower rate) and a small value of $m$ (so it reaches a larger acceleration). For the systems that we have considered in this work, we find $\eta \approx 10^{-13}$.

In summary, we have investigated the optical force produced by thermal radiation on a structure consisting of two planar layers of different materials. We have found that this force becomes maximal when one of the layers behaves as a perfect absorber and the other one as a perfect reflector. While this is just an ideal situation, we have shown that, using realistic materials, it is possible to obtain forces of the same order of magnitude. In our study, we have considered Ag as the reflecting layer, while we have analyzed different materials with high absorptivity in the infrared for the absorbing layer: C, SiC, SiO$_2$, and CaF$_2$. Our results show that, for a temperature difference of a few hundred degrees Kelvin between the structure and the environment, the force per unit area can reach values approaching $10^{-16}$~$\rm N/{\mu m}^2$. Furthermore, both the force and the resulting acceleration are highly dependent on the thickness of the absorbing layer. However, while the force always increases with thickness, the acceleration is maximized for a certain thickness as a result of the tradeoff between force and mass, reaching values as large as $10^4\,\mu$m/s$^2$ for structures with a thickness of $\approx 1\,\mu$m.
The motion of the planar structure is ultimately fueled by the thermal energy stored in the system, which is turned into kinetic energy. Consequently, the structure accelerates as it thermalizes with the environment, reaching terminal velocities up to $\approx 600\,\mu$m/s in a time scale of seconds for the largest temperatures under consideration. Besides the insights into optical forces produced by thermal radiation, our work explores a way to use them for the manipulation of material objects in the nano and microscale.

In nanoscale systems, the source of heat in the structure could originate from the physical contact with a hot material, such that motion would be produced once such contact is released. Heat could also be produced by nuclear decay in a radioactive layer separating the absorbing and reflective layers. In such a scenario, the transformation of the products of the radioactive decay into heat would require to also optimize the thickness of the structure to maximize heat deposition. Substantially larger terminal velocities could be reached in this configuration. For example, a layer of uranium with 20\% of U-235 would release an energy of $10^{17}$~J/m$^3$, orders of magnitude larger than the thermal energy corresponding to an initial temperature $T(0)$ of 1000s~K.

\begin{acknowledgements}
This work has been supported in part by projects PID2019-109502GA-I00, PID2022-137569NB-C42, PID2020-112625GB-I00, and Severo Ochoa CEX2019-000910-S funded by MCIN/AEI/10.13039/501100011033/FEDER, UE, the Catalan CERCA program, and Fundaci\'os Cellex and Mir-Puig.  J. R. D-R. acknowledges support from a predoctoral fellowship from the MCIN/AEI assigned to PID2019-109502GA-I00. 
\end{acknowledgements}


%

\end{document}